\documentclass[a4paper,fleqn,usenatbib,useAMS]{mnras}

\usepackage{graphicx}	
\usepackage{amsmath}	
\usepackage{amssymb}	
\usepackage{multicol}        
\usepackage{bm}		
\usepackage{pdflscape}	
\usepackage[T1]{fontenc}
\usepackage{ae,aecompl}
\usepackage{longtable}
\newcommand{\kms}{\,km\,s$^{-1}$} 

\title[The mid-IR spectrum of O stars]{Mid-infrared observations of O-type stars: spectral morphology
		\thanks{Based on observations with the NASA {\it Spitzer Space Telescope},
		which is operated by the Jet Propulsion  Laboratory (JPL), California Institute
		of Technology (Caltech) under a  contract with National Aeronautics and
		Space Administration (NASA).}}
\author[W. Marcolino]{W. L. F. Marcolino$^{1}$\thanks{Contact e-mail: \href{mailto:wagner@astro.ufrj.br}{wagner@astro.ufrj.br}}, 
J.-C. Bouret$^{2}$, T. Lanz$^{3}$, D. S. Maia$^{1}$, and M. Audard$^{4}$
\\
$^{1}$Universidade Federal do Rio de Janeiro, Observat\'orio do Valongo, Ladeira Pedro Ant\^onio, 43, Rio de Janeiro, Brasil\\
$^{2}$Aix Marseille Univ, CNRS, LAM, Laboratoire d'Astrophysique de Marseille, Marseille, France \\
$^{3}$Observatoire de la C\^ote d'Azur, Nice, France \\
$^{4}$Department of Astronomy, University of Geneva, ch. d'Ecogia 16, CH-1290 Versoix, Switzerland}
\date{Last updated 2017 April; in original form 2017 March}

\pubyear{2017}

\begin{document}
\label{firstpage}
\pagerange{\pageref{firstpage}--\pageref{lastpage}}
\maketitle

\begin{abstract}
We present mid-infrared observations for a sample of 16 O-type stars. The data
were acquired with the NASA {\it Spitzer Space Telescope}, using the IRS
instrument at moderate resolution (R $\sim$ 600), covering the range $\sim 10-37$
\micron.  Our sample includes early, mid and late O supergiants and dwarfs. We
explore for the first time their mid-IR spectral morphology in a quantitative way.
We use NLTE expanding atmosphere models to help with line identifications, analyze
profile contributions and line-formation regions. The O supergiants present a rich
emission line spectra. The most intense features are from hydrogen - $6\alpha$, 
7$\alpha$, and $8\alpha$ - which have non-negligible contributions of \ion{He}{I} 
or \ion{He}{II} lines, depending on the spectral type. The spectrum of early O 
supergiants is a composite of \ion{H}{i} and \ion{He}{ii} lines, \ion{He}{i} 
lines being absent. On the other hand, late O supergiants present features 
composed mainly by \ion{H}{i} and \ion{He}{i} lines. All emission lines 
are formed throughout the stellar wind. We found that O dwarfs exhibit 
a featureless mid-IR spectrum. Two stars of our sample exhibit very similar 
mid-IR features, despite having a very different optical spectral classification. 
The analysis of O-type stars based on mid-IR spectra alone to infer spectral 
classes or to estimate physical parameters may thus be prone to substantial errors. 
Our results may therefore inform spectroscopic observations of massive stars 
located in heavily obscured regions and help establish an initial 
framework for observations of massive stars using the Mid-Infrared Instrument 
on the {\it James Webb Space Telescope}.
\end{abstract}



\section{INTRODUCTION}

Massive O stars have been extensively studied in the last 
decades through the comparison of synthetic spectra with 
high-resolution spectroscopic data. Most of such investigations found 
in the literature focus on ground-based optical observations, and some also 
include satellite ultraviolet (UV) observations. Depending on the luminosity class, 
both the photosphere and wind of an O star can be detected 
in the optical (e.g., in early supergiants). However, the major energy output
from massive stars is in the UV region, where most stellar wind diagnostics can be found. 

\begin{table*}
  \caption{Log of the {\it Spitzer Space Telescope} mid-IR observations. Exposure times include target plus sky background subtraction.}
  \label{logobs}
  \begin{tabular}{lccccc}
    \hline
    Star               	& Spectral type 	& Total exposure time (s)  &  Total exposure time (s) & Obs. Date  & Note\\
      			&	              	&  IRS/SH                  & IRS/LH                   &       &        \\
    \hline 
    HD~66811 	       & O4If+    		& 38   			& 50    	& 2007-10-19 & $\zeta$ Puppis   \\  
                       &                        &                       &               & 2007-10-23 & \\
    HD~190429A         & O4If+   		& 218  			& 1935 		& 2004-06-26 & \\
    HD~188001          & O7.5Iaf  		& 244  			& 1935  	& 2006-11-14 & 9 Sge \\
                       &                        &                       &               & 2007-06-13 & \\
    HD~207198          & O8.5II  		& 126 			& 967   	& 2006-10-18 &  \\
    HD~30614 	       & O9.5Iab  		& 63   			& 117   	& 2006-10-18 & $\alpha$ Cam    \\
                       &                        &                       &               & 2007-10-03 & \\
    HD~188209          & O9.5Iab  		& 244  			& 1935  	& 2006-10-18 & \\
    HD~209975          & O9.5Ib   		& 126  			& 488   	& 2006-09-15 & 19 Cep \\
    HD~195592          & O9.7Ia  	        & 126 			& 244   	& 2006-10-18 &  \\
                       &                        &                       &               & 2006-11-13 & \\
    \hline
    HD~46223		& O4V((f)) 		& 488			& 3869		& 2007-04-27 &	 \\
    HD~46150 		& O5V((f)) 		& 488			& 2902		& 2007-04-27 & \\
    HD~199579           & O6V((f))  	        & 244  			& 967   	& 2007-06-18 & \\
    HD~206267           & O6.5V((f))		& 244  			& 488   	& 2006-10-18 & \\
    HD~47839            & O7V((f))   		& 488  			& 4837 		& 2007-04-27 & 15 Mon \\
                        &                       &                       &               & 2007-05-02 & \\
    HD~209481          & O9V  		        & 488 			& 3869 		& 2006-10-18 & LZ Cep \\
    HD~214680          & O9V      	        & 488  			& 4837 		& 2007-08-03 & 10 Lac  \\
                       &                        &                       &               & 2007-09-02 & \\
    HD~38666            & O9.5V      	        & 488			& 3869 		& 2007-10-06 & $\mu$ Col  \\
    \hline
  \end{tabular}
 \end{table*}

After the first rocket experiments (\citealt{Morton1967}) and with the early satellite missions in the UV 
(e.g., {\it Copernicus}, {\it IUE}), a variety of P-Cygni line profiles of different ions 
was revealed for the first time (e.g., \ion{N}{V} $\lambda\lambda$1239, 1243, \ion{O}{V} $\lambda$1371, \ion{Si}{IV} $\lambda\lambda$ 1394, 1403, 
\ion{C}{IV} $\lambda\lambda$1548,1551, \ion{He}{II} $\lambda$1640, and \ion{N}{IV} $\lambda$1718). 
It was soon discovered that the UV morphological properties of the O-type stars is
strongly correlated to their optical spectral classifications (\citealt{Walborn..Panek1984}). 
Since then, these lines were continuously used to infer the mass-loss rates and terminal wind 
velocities of O stars of various spectral types (\citealt{Howarth..Prinja1989}). In addition, detailed analyses 
of UV data revealed several new physical informations (e.g., NACs, superionization, clumping; 
\citealt{Prinja..Howarth1986}; \citealt{Bouret2005}) and provided ways to check parameters inferred from optical 
analyses (e.g., effective temperatures;  \citealt{Crowther2002}).

Compared to the UV and optical studies, there are relatively few high-resolution observations 
of O stars at longer wavelengths. In the infrared (IR), the energetic output of O-type 
stars is considerably weaker than in the optical and UV regions. Moreover, depending on the IR band, 
several observational difficulties appear (e.g., telluric contamination, sky and thermal emission).
Nonetheless, near and mid-IR data present obvious advantages in some specific contexts. 
For example, at various Galactic locations, the visual extinction is very large, 
reaching $A_V \sim 20-30$ (e.g., obscured H \ion{}{II} regions; Galactic center), and
the analysis and interpretation of J, H and K band spectra remains the best opportunity.
In this context, \cite{Lenorzer2004} were able to show that the \ion{He}{i-ii} line ratios 
in the near-IR correlate well with the line ratios in the optical spectrum that are used for spectral classification 
(see also \citealt{Hanson1996}). Soon after, \cite{Repolust2005} demonstrated for a sample of OB stars 
that photospheric and wind parameters derived from the near-IR alone matched well 
the ones derived from the optical (within error bars). That is, {\it a priori} one can obtain accurate 
physical parameters of OB stars which are not accessible at visual or UV wavelengths because
of high extinction.

However, we should note that a direct spectral 
classification or derivation of physical parameters based only on the infrared is not problem-free.  
For instance, \cite{Conti1995} found two Of stars with K band spectra 
typical of Wolf-Rayet stars (WN type), despite having normal optical and UV characteristics 
found in the Of class. By analyzing evolved massive stars, \cite{Morris1996} reported a 
similar problem, where objects from a certain group -WNL, Of, Of/WN, Be, B[e] or LBV - 
can be classified as a member of another class if one relies uniquely on the 2\micron\, spectra.
An analogous issue is pointed out again in the present paper.

Combined with other spectral regions, long wavelength data can provide valuable additional 
constraints on important physical parameters. In the near-IR for example, the Br$\alpha$ line 
emerged as an important mass-loss rate diagnostic in O stars (\citealt{Lenorzer2004}; 
\citealt{Najarro2011}). The IR and radio region have also long been known to provide 
mass-loss values through continuum flux excesses (\citealt{Wright..Barlow1975}). 
However, given the very low fluxes, radio band studies are usually restricted to the brightest objects. 
More recently, radio plus IR and optical data were used by \cite{Puls2006} to evidence
strong constraints on the radial stratification of wind clumping in O stars (density inhomogeneities).
Such multi-wavelength investigations are crucial to address the stellar and wind properties of massive stars. 
Uncertainties on stellar parameters can be reduced and model limitations might become clear
by using as many line and continuum diagnostics as possible. 

In this work, we present mid-IR observations of O stars and discuss their spectral 
morphology for the first time. The data were acquired with the IRS instrument on-board 
the NASA {\it Spitzer Space Telescope}. Our sample consists of 16 stars, comprising 
late, mid and early dwarfs and supergiants. We describe their mid-IR spectral morphology 
with the aid of NLTE stellar atmosphere models and complementary ultraviolet and optical data. 
We present line identifications, equivalent width measurements, profile formation 
regions and ion contributions. There are clear differences among 
the spectral types. O dwarfs present featureless atmospheric spectra. 
In supergiants, the mid-IR lines - essentially from hydrogen and helium - increase 
considerably in strength from late to early supergiants. Interestingly, two stars of 
our sample present very similar mid-IR spectra, despite having different optical 
spectra (spectral types).  In Sect.~\ref{obs}, we start by presenting our sample along 
with the observational details. In Sect.~\ref{morph}, we show the observed 
spectra and discuss the mid-IR spectral morphology in detail. Our main conclusions are 
summarized in Sect.~\ref{theend}.

\section{OBSERVATIONS}
\label{obs}

We used the IRS instrument onboard the NASA {\it Spitzer Space Telescope} to obtain 
mid-infrared high-resolution spectra for a sample of 16 O stars (PI T. Lanz, program 30030). 
The Short-High (SH) and Long-High (LH) modules were used (\citealt{Houck2004}). Both are 
cross-dispersed echelle spectrographs which provide a wavelength coverage 
of $\sim 9.9 - 19.6\mu$m and $\sim 18.7 - 37.2\mu$m, respectively. 
The detectors are $128 \times 128$ pixel Si:As blocked impurity band arrays, with 2.3 arcsec per pixel (SH) 
and 4.5 arcsec per pixel (LH). The exposure times were estimated to aim for signal-to-noise 
ratios of about 150 and 50 with the SH and LH modules, respectively. The data reduction was 
performed from the processed pipeline data, using post-BCD tools and the SPICE package (\citealt{SPICE}).
The final spectra were obtained by co-adding the different sub-exposures (2 to 4 individual 
exposures for SH and LH, respectively) and by merging the different orders with an IDL routine.
They were all normalized by polynomial fits to the continua, each specified by a careful selection 
of points in continuum windows. Our sample consists of 8 dwarfs and 8 supergiants, comprising late, mid and early types. 
The stars and corresponding spectral classes (\citealt{Sota2014}) are presented in Table~\ref{logobs}. 
Total exposure times for each module are indicated, including time used for sky background substraction. 
For HD~190429A, we did not repeat observations in the SH range and made 
use of previous IRS calibration observations, namely, from {\it AOR10021376}.

In order to achieve our goals, we also use  high-resolution 
ultraviolet and optical observations to further 
analyze the mid-infrared region quantitatively by means of sophisticated 
atmosphere models (see Section \ref{morph}). In the optical, we used public 
and own data from the FEROS (R $\sim$ 48000), NARVAL (Telescope Bernard-Lyot; 
R $\sim$ 65000), ESPADONS (CFHT; R $\sim$ 68000), and ELODIE (OHP; R $\sim$ 42000) 
spectrographs. In the ultraviolet, we focused on SWP spectra from the {\it International Ultraviolet Explorer} 
satellite. 

\section{MID-IR SPECTRAL MORPHOLOGY}
\label{morph}


\begin{figure*} 
\centering
 \makebox[\textwidth]{\includegraphics[trim=10mm 10mm 10mm 20mm,height=.55\paperheight,width=\paperwidth,angle=180]{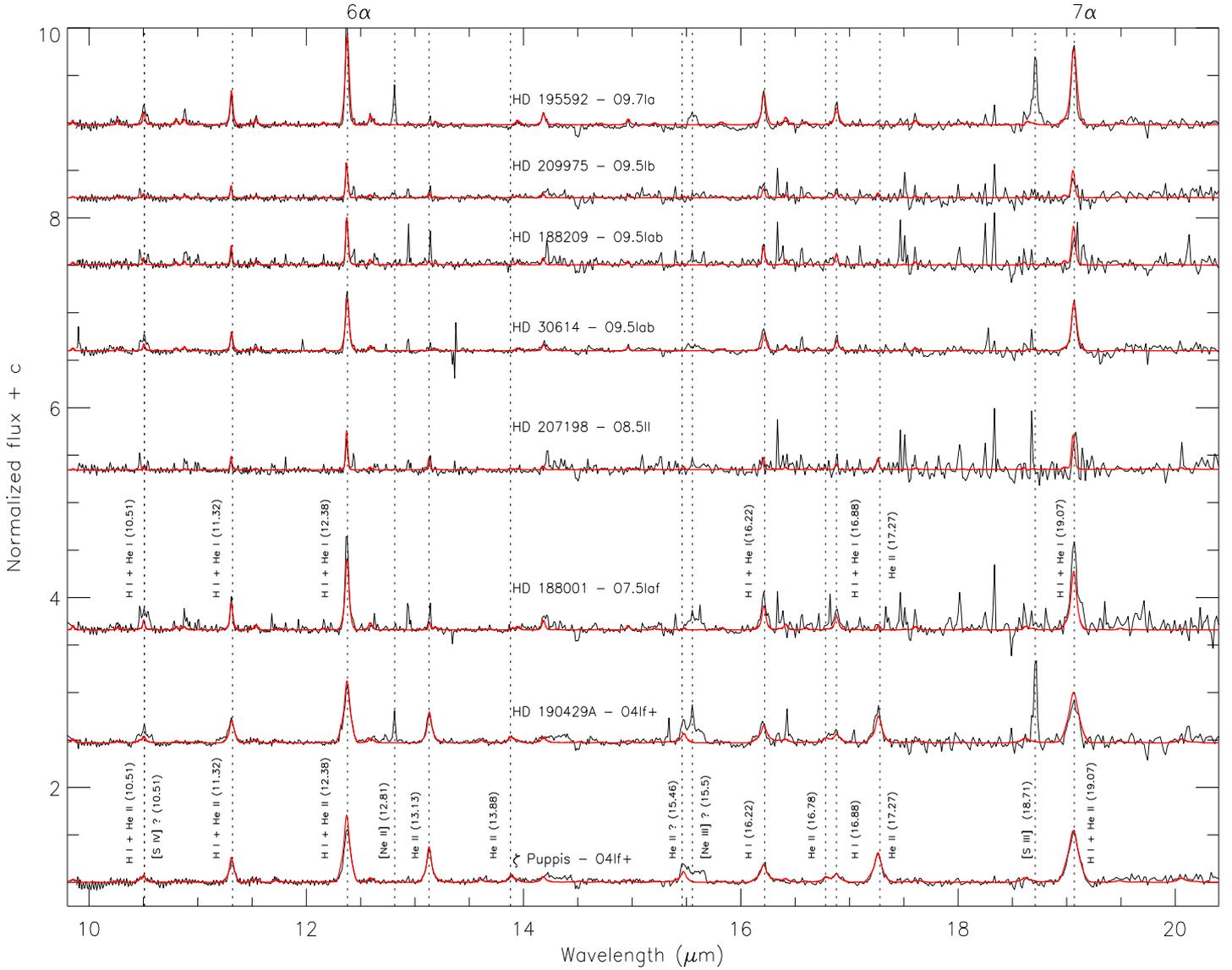}}
\caption{The mid-IR {\it Spitzer} spectra of the supergiants of our sample, between 9.8 and
20.4 \micron\, (black lines), with early-type to late-type supergiants from bottom to top.
Synthetic CMFGEN spectra are also displayed (red lines). They are essential
to characterize quantitatively the mid-IR spectrum, providing the line
identifications shown, individual ion contributions and line formation regions (see text for more details).}
\label{sgs1}
\end{figure*}

\begin{figure*} 
\centering
 \makebox[\textwidth]{\includegraphics[trim=10mm 10mm 10mm 20mm,height=.55\paperheight,width=\paperwidth,angle=180]{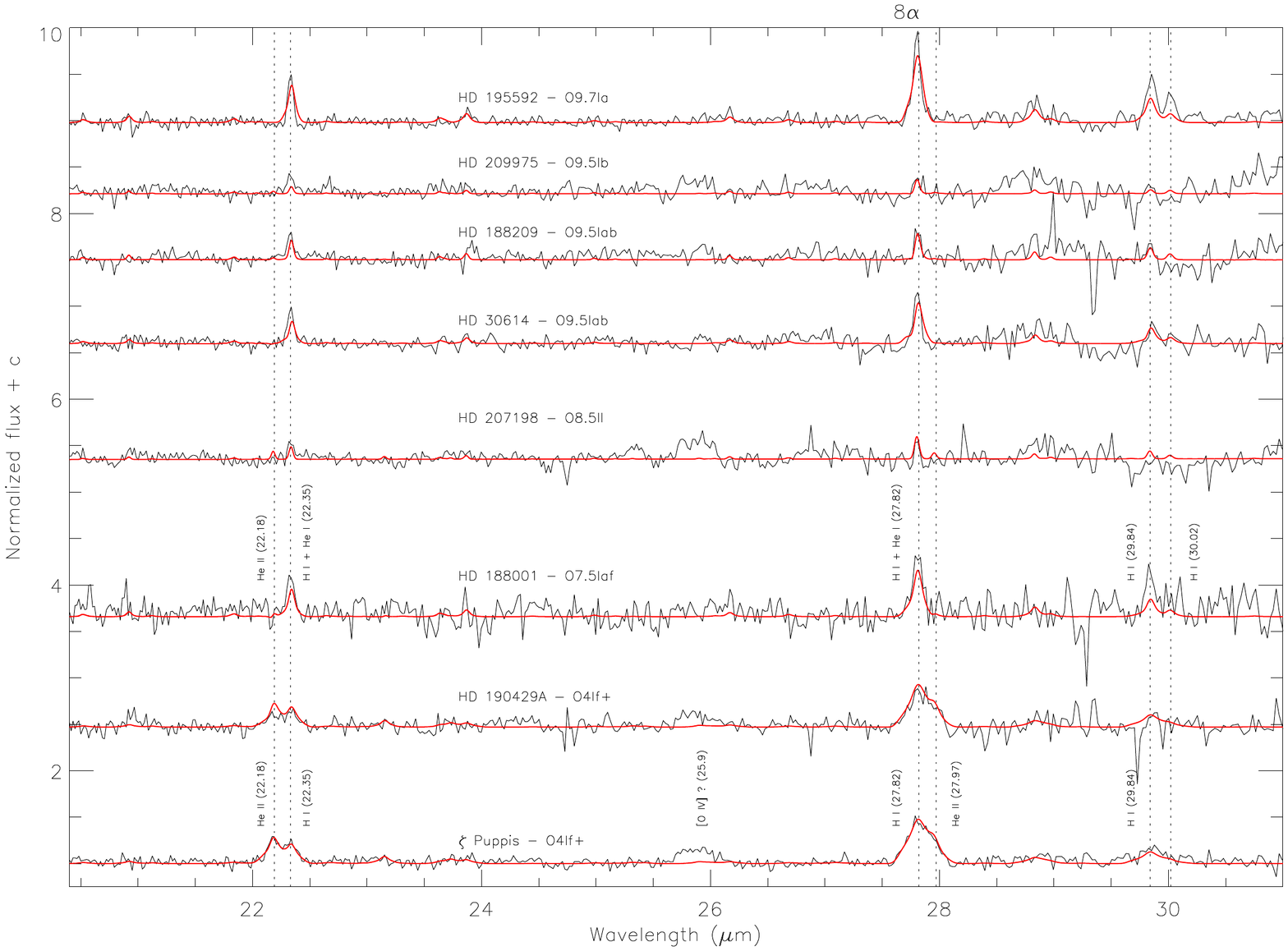}}
\caption{Same as Fig. \ref{sgs1} but for the spectral region between 20.4 and 31.0 \micron.}
\label{sgs2}
\end{figure*}

Previous mid-IR spectra of O-type stars were shown by \cite{Ardila2010} in their 
Spitzer Atlas of Stellar Spectra (SASS). The atlas contains a total of 159 objects of several 
spectroscopic types, out of which 12 objects are O-type stars. It is most complete at spectral types later than the G type, 
for luminosity classes V (dwarfs) and III (giants). The O stars are not analyzed in detail, neither 
qualitatively nor quantitatively. \cite{Ardila2010} only emphasized that the spectra contain 
emission lines that are typical of wind or circumstellar material. 
The resolving power of the SASS observations is $\sim 60-130$, with the low-resolution mode of the IRS,
that is a factor of about 5 to 10 lower than the resolution of the spectra shown here.
Previous observations with other space observatories (e.g., {\it ISO}, {\it IRAS}) are also 
of lower resolution, focusing mainly in spectral types and objects other than O stars 
(see e.g., the compilation by \citealt{Hodge2004}).

In order to provide a robust morphological analysis of the spectra, we relied on stellar atmosphere models
computed with the code CMFGEN (\citealt{CMFGEN}). We provide line identifications, equivalent width measurements ($W_\lambda$), 
compute individual ion contributions and line formation regions to fully characterize the mid-IR spectrum 
of O stars for the first time. We refer the reader to \cite{Bouret2012} for a detail description of the code 
and a canonical approach for a thorough analysis of massive stars. For several stars, we did not start the quantitative 
analysis from scratch. Initial models were taken from previous investigations by our group 
(\citealt{MartinsVAR2015}; \citealt{MartinsCNO2015};  \citealt{Bouret2012}). For completeness, we present the 
photospheric and wind parameters of the final models in Table \ref{params}. 

A detailed discussion about the models, their sensitivity to the parameters and the potential of mid-IR spectra to
constrain the physical properties, especially useful in the context of the arrival of JWST, will be presented in
a forthcoming paper. Here, we focus on the mid-IR morphology. We present and discuss the spectra of our sample below, 
starting with the O supergiant stars.

\subsection{O supergiants}

The spectra of the O supergiants of our sample are displayed in Figs. \ref{sgs1} and \ref{sgs2}. 
For clarity, they were normalized and offset by a constant. We cut data beyond $\sim 31$ \micron, 
due to very poor signal-to-noise ratio. 
 
The observed spectra are characterized by a wealth of emission lines. The
identification of these features was made more reliable thanks to the synthetic
spectra computed with CMFGEN, which agree well for the whole sample with observations throughout the
entire spectral range. We stress that this is truly
remarkable since the parameters of the models were constrained from the analysis
of the optical and UV spectra. Although models are not the focus of the
present paper, we illustrate this excellent agreement in Fig. \ref{finalzeta} using  the O4If+
star $\zeta$ Puppis as an example. The high fidelity of the synthetic spectra,
over such a broad wavelength range, is a major achievement of stellar atmosphere
modeling\footnote{We note that our best models initially predicted a relatively
intense \ion{H}{I} line at 13.19 \micron\, ($n = 18 \rightarrow 10$), which is
absent in all observed spectra. After some investigation, we found that the $f$
and $A$ values for this transition in the CMFGEN atomic database is higher than it
should be by a factor of 10, according to the {\it NIST} database (\citealt{NIST};
John Hillier, private comm.).}.  

The most intense spectral lines in the mid-IR spectrum of the O-type supergiants
correspond to wavelengths of the \ion{H}{i} lines: $6\alpha$ (12.37\micron),
$7\alpha$ (19.06\micron) and $8\alpha$ (27.80\micron). They arise from high-level
transitions: n $= 7 \rightarrow 6$, n $= 8 \rightarrow 7$, and n $= 9 \rightarrow
8$, respectively. However, as we will show below, these lines and in fact most
\ion{H}{i} transitions also contain a contribution of \ion{He}{i} or
\ion{He}{ii} lines, depending on the spectral type (or, more precisely, of the effective
temperature).  

The hottest two stars, with spectral types O4, also show a few
individual lines of \ion{He}{ii}, most intense at 13.13\micron\, (11-10), 17.26\micron\, (12-11),
22.17\micron\, (13-12), and significantly weaker at 15.47\micron\, (15-13), 16.77\micron\, (18-15). 
Note that the line at 17.26\micron\, is also observed in HD 207198, despite its much later
spectral type (O8.5). The use of models allowed to pinpoint this line, as it is
weak and not much stronger than the noise level in surrounding zones.   

\begin{figure}
\centering
\includegraphics[trim= 10mm 10mm 10mm 10mm,width=8.5cm,height=7.5cm,angle=180]{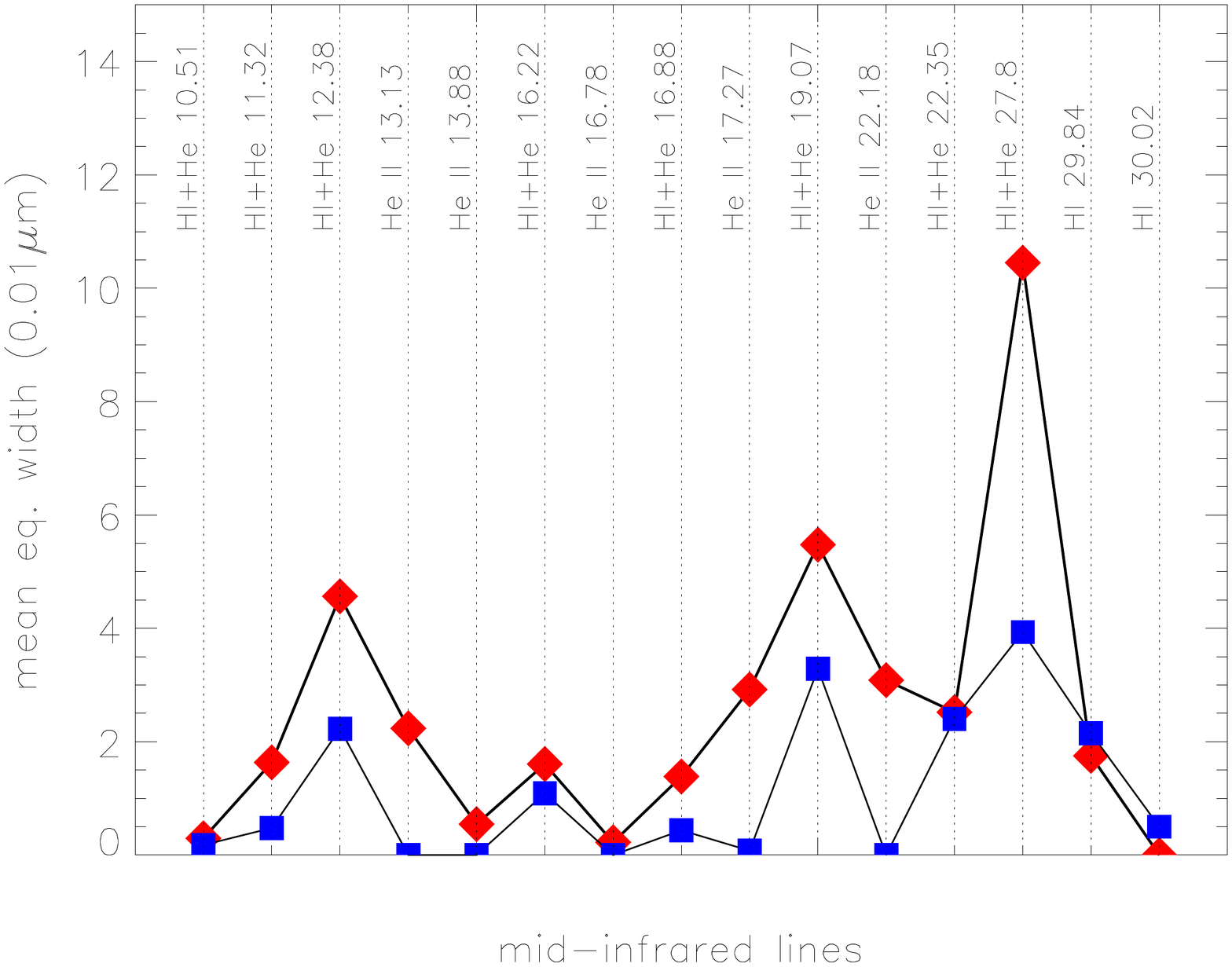}
\caption{Mean equivalent widths of mid plus late-type O supergiants ($\ge$ O7; squares; thin
solid line) and early-type O supergiants (diamonds; thick solid line). Note that
the highest values correspond to the wavelengths of the 6$\alpha$, 7$\alpha$, and
8$\alpha$ transitions. The \ion{He}{ii} transitions at 13.13\micron, 17.27\micron,
and 22.18\micron, clearly separate early-type objects from later-type stars.}
\label{ew}
\end{figure}

We present a detailed list of line identifications and W$_{\lambda}$ measurements
in Table \ref{EWtable}.  In some cases, no reliable W$_{\lambda}$ could be
obtained (e.g., weak or blended feature). Therefore, the respective line
identifications on the table should be then interpreted as an indication that a
line is likely present, based on the overlying models. These results are
summarized in graphical form in Fig. \ref{ew}, where we present mean equivalent
widths of the late plus mid ($\ge$ O7) and earlier supergiants classes. The
W$_{\lambda}$ values increase from late to the early supergiants. Its clear that
early supergiants have the most intense lines. Moreover, regardless of the spectral class, the
most intense features are always $6\alpha$, $7\alpha$, and $8\alpha$. 

We did not identify metallic lines (e.g., from CNO ions) in a definitive way, in contrast
to spectral features observed in the near-IR. For example, \cite{Hanson1996} report lines of
\ion{C}{iv} and \ion{N}{iii} in several O stars spectra around 2\micron. The lines
that are not accounted for by our model spectra are unlikely to be formed in
the atmosphere of the stars. This is for instance the case of the features between
$\sim 16-19$\micron, most clearly visible in the mid-to-late supergiants, and for
which we could not find any identification in the literature (see Fig.~\ref{sgs1}).
We do not exclude the possibility that some of them might be
artifacts that remained despite the data reduction process. The features at $\sim
15.5$\micron\, and close to 26\micron\, (box-like shape) are also
clear discrepancies. They are both conspicuous in $\zeta$ Puppis and HD~190429A (see
Figs.~\ref{sgs1} and \ref{sgs2}). They were also reported in the {\it Spitzer}
observations of EZ Canis Majoris, a WN star, by \cite{Morris2004} who identified
them to [\ion{Ne}{iii}] 15.5\micron\, and [\ion{O}{iv}] 25.9 \micron\, transitions. 

The [\ion{S}{iv}] 10.5 \micron, [\ion{Ne}{ii}] 12.8 \micron, and [\ion{S}{iii}] 18.7\micron\, lines, are also
possibly present in some spectra, e.g. in HD~190429A and HD~195592. These two stars are 
known binaries. The respective companions of each were certainly in the field of view of 
the SH and LH modules during observations, but it remains speculative to assume that the companions are responsible 
for the observed features. We note that HD~195592 has a very high reddening (E(B-V) = 1.17, 
see \citealt{McSwain2007}), indicating that lots of interstellar/circumstellar 
material is on the line-of-sight. Hence we are tempted to assume that the emission features are produced around 
HD~195592. Interpreting their presence in HD~190429A is more difficult as E(B-V) = 0.46 (\citealt{Bouret2012}) 
and is nothing exceptional even for early-type supergiants. In any case, such fine-structure transitions 
must be formed far from the bulk of the stellar wind, where densities are very low (log $n_e$/cm$^3$ $\sim 5$). 

Despite our efforts, several non-atmospheric features remained unidentified. We 
note that the signal-to-noise starts to deteriorate beyond 28\micron\, and it becomes hard to confirm the 
presence of lines at longer wavelengths, regardless the origin.

\begin{figure}
\centering
\includegraphics[trim= 25mm 10mm 20mm 0mm,width=8.5cm,height=9cm,angle=180]{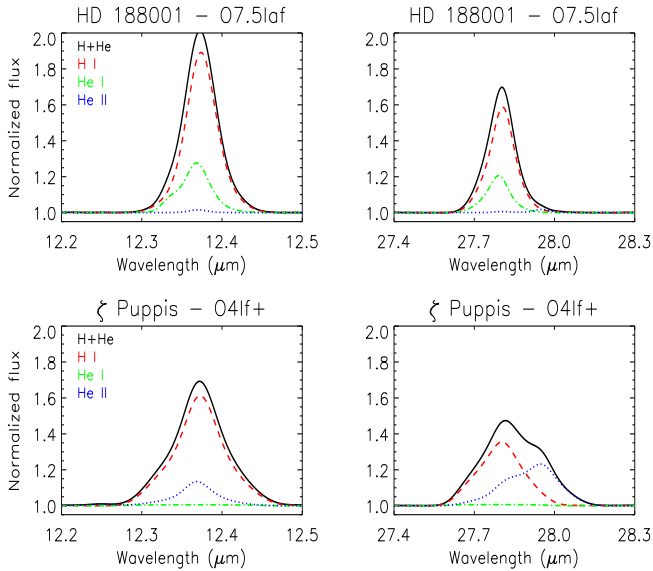}
\caption{Individual ion contributions to the most intense mid-IR profiles in O
supergiants. Top panels correspond to HD~188001 (O7.5Iaf). Bottom panels
correspond to $\zeta$ Puppis (O4If+). For clarity, only the synthetic lines
6$\alpha$ and 8$\alpha$ are shown (full spectrum: black full line; \ion{H}{i}: red
dashed line; \ion{He}{i}: green dashed-dotted line; \ion{He}{ii}: blue dotted
line).}
\label{contributions}
\end{figure}

\begin{figure}
\centering
\includegraphics[trim= 20mm 10mm 20mm 10mm,width=8.5cm,height=8cm,angle=180]{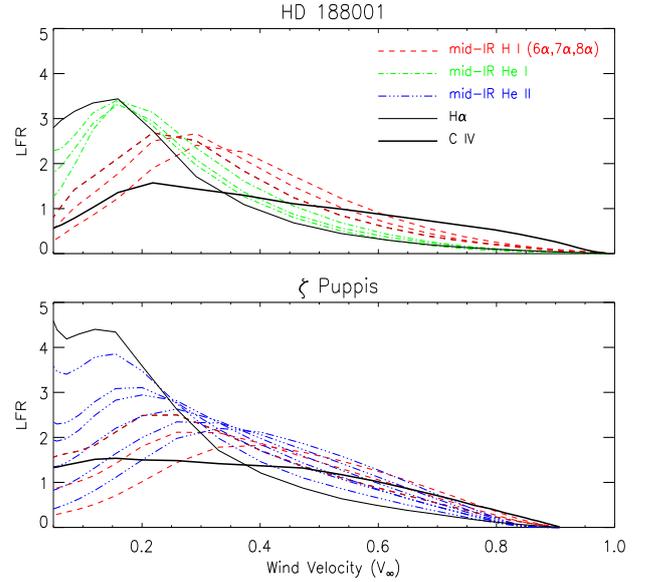}
\caption{Line formation regions in O supergiants. The \ion{H}{i} transitions are the 
mid-IR  (6$\alpha$, 7$\alpha$, and 8$\alpha$) and H$\alpha$. The \ion{He}{i} and 
\ion{He}{ii} transitions chosen are at $\sim$12.36, 19.06, and 27.9 \micron, 
i.e., at wavelengths corresponding to 6$\alpha$, 7$\alpha$, and 8$\alpha$. \ion{He}{ii} transitions 
at 13.13, 17.26, and 22.17 \micron\, are also shown.}
\label{LFRs}
\end{figure}

\begin{figure*}
\centering
\includegraphics[trim= 20mm 20mm 20mm 20mm,width=16cm,height=14cm,angle=180]{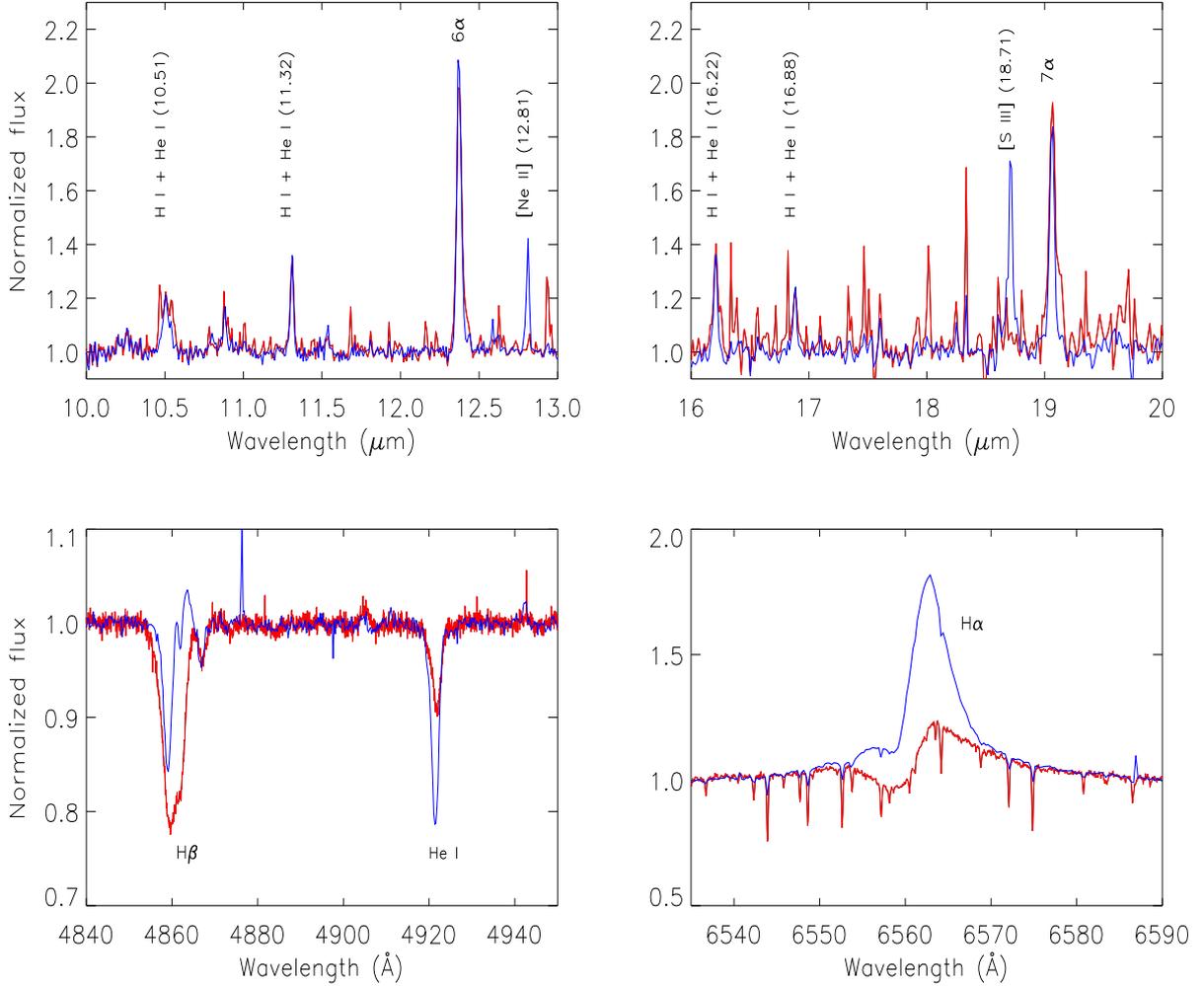}
\caption{The similar mid-IR (top) and drastically different optical spectra
(bottom) of HD~195592 (O9.7Ia, blue lines) and HD~188001 (O7.5Iaf, red lines). Note that the mid-IR mismatches -
in particular around 16-20\micron\, - are likely due to non-atmospheric features, as
discussed in Section \ref{morph}.}
\label{twosuper}
\end{figure*}

\subsubsection{Line contributions and formation regions}

Following our methodology to identify the transitions in the spectra\footnote{We
have computed synthetic spectra produced by individual ions and checked the
corresponding wavelengths.}, we found that the profiles 6$\alpha$, 7$\alpha$,
and 8$\alpha$ (the most intense ones) are contaminated by a \ion{He}{i-ii}
contribution, as illustrated in Fig.~\ref{contributions}. For simplicity, we
selected regions comprising the 6$\alpha$ and 8$\alpha$\, lines and show only
synthetic models for HD~188001 (O7.5Iaf) and $\zeta$ Puppis (O4If+) as examples.
In early O supergiants, the full line profiles are formed by a contribution of
\ion{H}{i} and \ion{He}{ii}. The \ion{He}{i} contribution is negligible. The
\ion{He}{ii} contribution to 8$\alpha$ is particularly important, and is
responsible for the asymmetry of the line profile. In mid and late supergiants, we see
the opposite: the profiles mostly result from the \ion{H}{i}\, transition, with a small contribution
of \ion{He}{i}, and \ion{He}{ii} is absent. {\it These characteristics of the mid-IR
line spectra were more clearly evidenced thanks to the stellar atmosphere models}. 

Overall, from the observations and model spectra, we conclude that: (i) the early
supergiants have the most intense transitions in the mid-IR; (ii) mid and late
supergiants present the main features composed by \ion{H}{i} and \ion{He}{i} lines;
(iii) early supergiants present the main features composed by \ion{H}{i} and
\ion{He}{ii} lines, \ion{He}{i} lines being absent; (iv) the most intense transitions
are from \ion{H}{i} $6\alpha$, $7\alpha$, and $8\alpha$.

Now, we present the line formation regions (LFRs) for specific lines. Regarding 
\ion{H}{i} lines, we chose $6\alpha$, $7\alpha$, and $8\alpha$. 
The \ion{He}{i} and \ion{He}{ii} transitions chosen are at $\sim$12.36, 19.05, and 27.9 \micron.
Besides mid-IR transitions, for comparison purposes, we compute LFRs for the 
\ion{C}{iv} $\lambda 1551$ and H$\alpha$ profiles. We refer the reader 
to \cite{Hillier1987} for an expression for the quantity that defines the LFR.

For simplicity, we present LFRs in Fig.~\ref{LFRs} for two stars, again 
HD~188001 (O7.5Iaf) and $\zeta$ Puppis (O4If+). It can be seen that $6\alpha$, $7\alpha$, 
and $8\alpha$ follows the \ion{C}{iv} ultraviolet line trend, being formed throughout the wind. 
The bulk of the \ion{He}{i} lines (in mid and late-type supergiants) is formed in the inner wind, 
similar to H$\alpha$. The \ion{He}{ii} LFRs (in early-type supergiants) follow more or less the ones 
of the mid-IR \ion{H}{i} lines.

In conclusion, {\it our results demonstrate that the observed mid-IR features in O supergiants are 
definitely formed in the stellar wind. Therefore, they are potential diagnostics for the determination of 
wind properties.} A detailed analysis of mid-IR line responses to the different physical parameters is beyond 
the scope of the present paper and will be presented elsewhere. Here, we just emphasize that 
the same value for the mass-loss rate (and clumping parameters) allows us to match simultaneously the UV P-Cygni profiles,
H$\alpha$, and the mid-IR lines, as demonstrated
for $\zeta$ Puppis (appendix, Fig. \ref{mdots}).

\subsubsection{The case of the two supergiants HD~195592 and HD~188001}

From a pure morphological, mid-IR point of view, HD~195592 is at odds with its
spectral classification, namely, O9.7 Ia. While its fellow late-type O9 (and
sub-division) supergiants exhibit weaker lines than the earlier type ones, the
lines in HD~195592 are remarkably similar in shape and intensity to those of
HD~188001, an O7.5 Iaf. This is best illustrated in Fig. \ref{twosuper}, where we
overplot a selection of lines from the mid-IR spectra of the two stars (but see
also Figs. \ref{sgs1} and \ref{sgs2} for a pan view). We also show portions of the
optical spectra, demonstrating that these stars are very different one of the
other, in agreement to their spectral classification. 
 
A low-resolution UV (IUE) spectrum for HD~195592 is available and despite the
modest signal-to-noise ratio, exhibits spectral morphology consistent with the
star's O9.7 Ia spectral type. So we are facing here an interesting conundrum, and taken at
face value, these two stars demonstrate that the mid-IR range should be used with
caution to assign a spectral type to a massive supergiant star, at least in the
range from O7.5 to O9.7 (where most supergiants are expected to belong). This is
specially important in the context of observations of heavily obscured regions.

Despite this seemingly paradoxical behavior, models can be found that
simultaneously fit the UV, optical, and mid-IR spectra. The parameters from these
models can be found in Table~\ref{params}. The model parameters are fully compatible
with the results found in the literature for stars of these spectral types, hence clearly
suggesting a spectral degeneracy in the mid-IR (different stellar parameters but very similar
spectra). This issue deserves a deeper theoretical investigation that is beyond the
scope of this paper.

In contrast to the arguments presented above, one can argue that (drastic) variability (O9 to O7) is a 
possible explanation for the atypical spectrum of HD~195592. Variability of HD~195592 from an "normal" 
mid-IR O9 spectrum (i.e., like the others O9 in the sample) to an O7 spectrum (matching HD~188001) 
and back (if periodic) would be drastic, but not impossible. Binarity effects, wind instabilities, 
and wind collisions are all possibilities to produce spectral changes. If this is the case,  
our observations would then correspond to HD~195592 at the spectral state of an O7 (HD~188001).
We note however that the SH and LH spectra of HD~195592 (and of HD~188001) were acquired 
at different dates (see Table \ref{logobs}). That is, in Figs. \ref{sgs1} and \ref{sgs2} we actually have 
four spectra. If variability is very important, similar spectra would be unlikely.

\subsection{O dwarfs}

The mid-IR spectra of O dwarfs - late, mid and early types - are featureless, that
is, with an absence of conspicuous lines formed in the photosphere or in the
stellar wind. For completeness, they are presented in Fig.~\ref{dwfs}
along with our model stellar atmosphere spectra, adopted from an UV/optical analysis. 

The spectrum of HD~38666 has a very low signal-to-noise ratio at wavelengths longer than 15\micron, 
hindering the analysis. The 10 Lac spectrum is of better quality, but it does not show 
either stellar or wind transitions. In HD~209481, the atmospheric transitions 
6$\alpha$, 7$\alpha$ and 8$\alpha$ are barely visible, if present at all. We recall that they 
are the strongest features in the spectra of O supergiants. Forbidden 
lines seem to be present in a few stars (e.g., [\ion{O}{iv}]) and various narrow 
unidentified transitions are conspicuous.

Although more luminous, hotter, and with higher mass-loss rates than O9V stars, 
the O4-7V stars of our sample also present a featureless spectra. 
Again, various (non-atmospheric) features at $\sim$ 17-19\micron\, and 
$\sim$ 23-27\micron\, (see e.g, HD~199579), as well as 
[\ion{O}{IV}] 25.9\micron\, (see e.g., HD~206267) are present.  

The results concerning the early dwarfs (O4-5V) are somewhat counterintuitive. 
Since their ultraviolet spectra contain intense wind emission lines 
- e.g., \ion{C}{IV} $\lambda\lambda$1548,1551 can be even found saturated, as in supergiants - 
one would expect at least weak emissions in the mid-IR. In fact, their effective 
temperature and luminosity values are even compatible with the ones found in early 
supergiants (see Table \ref{params}). We must remind however that their surface gravities
are higher by up to 1~dex relative to supergiants and most importantly, the mass-loss rates of the O 
supergiants can be higher by 1-2~dex. Therefore, it seems that a combination 
of low gravity and intense mass-loss is needed for a detectable mid-IR line 
emission spectrum. Nevertheless, a firmer conclusion awaits a more detail theoretical study.

Overall, {\it we conclude that there is no atmospheric information 
regarding O dwarfs in the mid-IR beyond the continuum flux level}. 
The observed features are likely formed far from their atmospheres, being 
circumstellar or even from the ISM (e.g., diffuse emission). Some are certainly spurious, 
given the limited signal-to-noise achieved for some stars (see e.g., HD~38666 and HD~199579 at $\sim 29\mu$m). 
Our models give confidence to this conclusion since they reveal pure mid-IR continua. 
This finding should be taken into account in future investigations. For example, it does not 
appear useful to acquire mid-IR spectra of main-sequence massive stars in regions
with extremely high visual extinction at high cost with the MIRI instrument 
aboard the JWST.

\section{CONCLUSIONS}
\label{theend}

We have presented an analysis of mid-IR data of O-type stars, based on observations with 
the NASA {\it Spitzer Space Telescope} (IRS instrument; R $\sim$ 600). We provided a 
detailed description of the spectra supported by model spectra from NLTE expanding model atmospheres 
calculated with the CMFGEN code. 
 
We found that O supergiants present a rich emission line mid-IR spectrum, while 
O dwarfs are essentially featureless in terms of atmospheric information.
The most intense transitions found in O supergiants are the \ion{H}{I} recombination lines: 6$\alpha$ 
($n = 7 \rightarrow 6$; 12.37\micron), 7$\alpha$ ($n = 8 \rightarrow 7$; 19.06\micron), 
and 8$\alpha$ ($n = 9 \rightarrow 8$; 27.80\micron). However, according to our models, 
these lines have a non-negligible contribution of \ion{He}{I} lines in mid and late O supergiants, and 
of \ion{He}{II} lines in early O supergiants. In early O-type supergiants, the \ion{He}{II} line contribution 
is responsible for the asymmetry of the $8\alpha$\, line profile. 

We provided line identifications and measurements for all detected \ion{H}{I}, \ion{He}{I}, and \ion{He}{II} 
features (Table \ref{EWtable}). The presence of isolated, relatively strong \ion{He}{ii} lines 
at 13.13\micron, 17.27\micron, and 22.18\micron, is a basic characteristic 
of early-type supergiant spectra. We do not identify lines from heavy element ions (e.g., CNO) formed in the
stellar photosphere or wind, in contrast to the near-IR spectrum. There are however several features 
that are probably of circumstellar origin, e.g., [\ion{Ne}{III}] 15.5\micron\, and [\ion{O}{IV}] 25.9 \micron.  
Despite our efforts, a few detected emission features remain unidentified.

Two supergiants -- HD~188001 and HD~195592 -- raised an interesting problem.
Their mid-IR spectra are very similar while their optical characteristics drastically differ. 
From a pure observational point of view, such fact means that the use of mid-IR  
data alone to classify O stars may be prone to errors if unsupported by observations in UV, optical or near-IR.
This must be taken into account in observations of heavily-obscured regions (e.g., the Galactic Center, 
obscured \ion{H}{II} regions). Additionally, the lack of photospheric or wind features in the mid-IR
spectrum of all O dwarfs put a stringent limitation about the usefulness of such observations.
In regions inaccessible to UV or optical observations because of heavy extinction, 
the near-IR should be preferred to observe and analyze main-sequence massive stars.

\section*{Acknowledgements}
\addcontentsline{toc}{section}{Acknowledgements}

The authors would like to thank D.~J.~Hillier for his continuous support 
regarding CMFGEN. WM would like to thank J.-C.~Bouret for 
the hospitality at the Laboratoire d'Astrophysique de Marseille, where 
part of the work was carried out. 
Support for this work was provided by NASA through an award issued by JPL/Caltech.
This research also made use of the SIMBAD database and Vizier service, 
operated at CDS, Strasbourg, France.

\bibliographystyle{mnras}
\bibliography{references} 



\clearpage
\appendix
\section{Additional data}

\begin{table*}
  \caption{Stellar and wind parameters of the sample. Values for additional physical (and chemical) parameters can be found in the quoted 
references: $(1)$ this work; $(2)$ Martins et al. (2015a); $(3)$ Martins et al. (2015b); $(4)$ Bouret et al. (2012). The uncertainty for T$_{eff}$ is 
about $\pm 2000K$, log $g$ is $\pm 0.1$, v$_{\infty}$ is about $\sim 200$ km s$^{-1}$, and $\dot{M}$ is $\sim$ 0.4 dex.}
  \label{params}
  \begin{tabular}{lccccccccc}
    \hline
    Star                      & Spec. type & T$_{eff}$  & log $g$     & L$_{\star}$      & log $\dot{M}$        & f$_{cl}$  & V$_{\infty}$  & $\beta$ & Ref.  \\
                              &            & (kK)      & (cgs)       & (L$_{\sun}$)     & (M$_{\sun}$ yr$^{-1}$) &          &  (\kms)     &         &       \\
    \hline 
    HD~66811                  & O4If+      & 41.0      & 3.6         & 5.92  $\pm$ 0.10  & -5.70            & 0.05    & 2300          &  0.9     & 1,4   \\ 
    HD~190429A               & O4If+      & 39.0      & 3.6         & 5.97  $\pm$ 0.10  & -5.64              & 0.04   & 2300         &   1.0     & 1,4   \\
    HD~188001                 & O7.5Iaf    & 33.0     & 3.4         & 5.69   $\pm$ 0.20  & -5.88             & 0.05   & 1800         &   1.5     & 1     \\
    HD~207198                 & O8.5II     & 32.5      & 3.5         & 5.05  $\pm$ 0.26  & -7.00              & 1.0   & 2000         &   3.0     & 1,2  \\
    HD~30614                   & O9.5Iab    & 28.9      & 3.0         & 5.81  $\pm$ 0.25  & -5.62              & 0.1     & 1600      &   1.2     & 1 \\
    HD~188209                 & O9.5Iab    & 29.8      & 3.2         & 5.65  $\pm$ 0.26  & -6.40              & 0.05    & 2000       &   2.2     &  1,2 \\
    HD~209975                 & O9.5Ib     & 30.5      & 3.3         & 5.35  $\pm$ 0.30  & -6.50              & 1.0     & 2000       &   2.9     & 1,2 \\
    HD~195592                 & O9.7Ia     & 28.0      & 2.9         & 5.47  $\pm$ 0.25  &  -5.79             & 0.05    & 1400        &  1.4     & 1  \\ 
    \hline
    HD~46223                   & O4V((f))  & 43        & 4.0	     &  5.60  $\pm$ 0.25   &   -7.20 		& 0.1     & 2800      &  1.0     &  3   \\
    HD~46150                    & O5V((f))  & 42       &  4.0        &  5.60   $\pm$ 0.25  &  -7.30  		 &  0.1    &  2800    &  1.0     &   3 \\
    HD~199579                 & O6V((f))   & 41.5      & 4.15        & 5.33 $\pm$ 0.25   &  -7.84             & 0.1     & 2750        &  0.8     & 1,3  \\ 
    HD~206267                 & O6.5V((f)) & 39.0      & 4.0         & 5.20 $\pm$ 0.25   &  -7.49             & 0.5     & 2750        &  0.8     & 1    \\ 
    HD~47839                  & O7V((f))   & 37.0      & 4.0         & 5.15 $\pm$ 0.25   &  -8.00             & 1.0     & 2000        &  1.0     & 1   \\ 
    HD~209481                 & O9V        & 31.5      & 3.9         & 4.72 $\pm$ 0.25   &  -8.70             & 1.0     & 1600        &  0.8     & 1  \\ 
    HD~214680                 & O9V        & 35.0      & 4.0         & 4.34 $\pm$ 0.25   & -9.52              & 1.0     & 1200        &  0.8     & 1,3 \\
    HD~38666                  & O9.5V      & 33.0      & 4.0         & 4.68 $\pm$ 0.25   & -9.82              & 1.0     & 1200        &  1.0     & 1,3 \\
    \hline
   \end{tabular}
 
\end{table*}

\begin{center}
  \begin{table*}
    \caption{Line identifications and equivalent width measurements of the mid-IR features of the O supergiants.}               
    \label{EWtable} 
    \begin{tabular}{ccccc}
      \hline \hline		
      Star  & Observed $\lambda (\mu m)$ & Identification & Rest Wavelength $\lambda (\mu m)$ & W$_\lambda$ ($\mu m \cdot 10^{-2}$) \\  \hline
      HD~66811 ($\zeta$ Pup)  &  &  &  &    \\ 
      & 10.51 & H I + He II         & 10.50 + 10.47 &  0.59 $\pm$ 0.01 \\ 
      & 11.32 & H I + He II         & 11.31 + 11.33 & 1.28 $\pm$ 0.04 \\ 
      & 12.38 & H I + He II         & 12.37 + 12.38 & 4.50 $\pm$ 0.07 \\ 
      & 13.13 & He II               & 13.13         &  2.33 $\pm$ 0.04 \\ 
      & 13.88 & He II               & 13.87         &  0.52 $\pm$ 0.04 \\ 
      & 15.46 & He II               & 15.47         &  -               \\ 
      & 16.22 & H I                 & 16.21         &  1.94 $\pm$ 0.03 \\ 
      & 16.78 & He II               & 16.77         & 0.45 $\pm$ 0.03 \\ 
      & 16.88 & H I                 & 16.88         &  1.26 $\pm$ 0.09 \\ 
      & 17.27 & He II               & 17.26         &  2.82 $\pm$ 0.07 \\ 
      & 19.07 & H I + He II         & 19.06 + 19.05 &  6.30 $\pm$ 0.04 \\ 
      & 22.18 & He II               & 22.17         & 3.44 $\pm$ 0.10 \\ 
      & 22.35 & H I                 & 22.33         &  3.15 $\pm$ 0.12 \\ 
      & 27.82 + 27.97 & H I + He II & 27.80 + 27.95 &  12.28 $\pm$ 0.11 \\ 
      & 29.84 & H I                 & 29.84         &  3.50  $\pm$ 0.50 \\ \hline
      HD~190429A &  &  &  &    \\ 
      & 10.51 & H I + He II      & 10.50 + 10.47 & -   \\ 
      & 11.32 & H I + He II      & 11.31 + 11.33 &  1.99 $\pm$ 0.12 \\ 
      & 12.38 & H I + He II      & 12.37 + 12.38 &  4.63 $\pm$ 0.07 \\ 
      & 13.13 & He II            & 13.13         &  2.14 $\pm$ 0.03 \\ 
      & 13.88 & He II            & 13.87         &  0.57 $\pm$ 0.02 \\ 
      & 15.46 & He II            & 15.47         &  -  \\ 
      & 16.22 & H I              & 16.21         &  1.27 $\pm$ 0.03 \\ 
      & 16.88 & He II + H I      & 16.77 + 16.88 &  1.51 $\pm$ 0.03 \\ 
      & 17.28 & He II            & 17.26         &  3.02 $\pm$ 0.03 \\ 
      & 19.08 & H I + He II      & 19.06 + 19.05 &  4.65 $\pm$ 0.16 \\ 
      & 22.19 & He II            & 22.17         &  2.73 $\pm$ 0.04 \\ 
      & 22.33 & H I              & 22.33         &  1.89 $\pm$ 0.23 \\ 
      & 27.82 & H I + He II      & 27.80 + 27.95 &  8.62 $\pm$ 0.13 \\ 
      & 29.84 & H I              & 29.84         &  -  \\ \hline
      HD~188001 &  &  &  &    \\ 
      & 10.51 & H I + He I    & 10.50         &  -  \\ 
      & 11.32 & H I + He I    & 11.31 + 11.30 &  1.05 $\pm$ 0.02 \\ 
      & 12.38 & H I + He I    & 12.37 + 12.39 &  4.55 $\pm$ 0.01 \\ 
      & 16.22 & H I + He I    & 16.21 + 16.20 &  2.38 $\pm$ 0.04 \\ 
      & 16.88 & H I + He I    & 16.88         &  1.10 $\pm$ 0.20 \\ 
      & 19.07 & H I + He I    & 19.06 + 19.05 &  7.82 $\pm$ 0.03 \\ 
      & 22.35 & H I + He I    & 22.33 + 22.32 &  3.88 $\pm$ 0.07 \\ 
      & 27.81 & H I + He I    & 27.80 + 27.79 &  7.40 $\pm$ 0.20 \\
      & 29.84 & H I           & 29.84         &  5.00 $\pm$ 0.30 \\ \hline     
      HD~207198 &  &  &  &    \\ 
      & 11.32 & H I          & 11.31         &  0.14 $\pm$ 0.01 \\ 
      & 12.38 & H I + He I   & 12.37 + 12.39 &  0.81 $\pm$ 0.02 \\ 
      & 17.27 & He II        & 17.26         &  0.46 $\pm$ 0.01 \\ 
      & 19.07 & H I + He I   & 19.06 + 19.05 &  1.49 $\pm$ 0.10 \\ 
      & 22.19 & He II   & 22.17 & - \\ 
      & 22.35 & H I + He I   & 22.33 + 22.32 &  1.29 $\pm$ 0.04 \\ 
      & 27.82 & H I + He I   & 27.80 + 27.79 &  0.65 $\pm$ 0.18 \\ \hline \hline
    \end{tabular}
  \end{table*}
\end{center}

\begin{center}
  \begin{table*}
    \contcaption{}
    \label{ap:continued} 
    \begin{tabular}{ccccc}
      \hline \hline		
      Star & Observed $\lambda (\mu m)$ & Identification & Rest Wavelength $\lambda (\mu m)$ & W$_\lambda$ ($\mu m \cdot 10^{-2}$) \\  \hline
      HD~30614 ($\alpha$ Cam) &  &  &  &   \\ 
      & 10.51 & H I + He I    & 10.50         & -               \\
      & 11.32 & H I + He I    & 11.31 + 11.30 & 0.45 $\pm$ 0.01 \\ 
      & 12.38 & H I + He I    & 12.37 + 12.39 & 2.56 $\pm$ 0.05 \\ 
      & 16.22 & H I + He I    & 16.21 + 16.20 & 1.32 $\pm$ 0.03 \\ 
      & 16.88 & H I + He I    & 16.88 + 16.87 & 0.55 $\pm$ 0.01 \\ 
      & 19.07 & H I + He I    & 19.06 + 19.05 & 3.07 $\pm$ 0.10 \\ 
      & 22.35 & H I + He I    & 22.33 + 22.32 & 2.56 $\pm$ 0.04 \\ 
      & 27.82 & H I + He I    & 27.80 + 27.79 & 4.17 $\pm$ 0.04 \\ 
      & 29.84 & H I           & 29.84         & 1.70 $\pm$ 0.20 \\
      & 30.02 & H I           & 30.02         & -                \\ \hline 
       HD~188209 &  &  &  &    \\ 
      & 10.51 & H I + He I   & 10.50         & -               \\
      & 11.32 & H I + He I   & 11.31 + 11.30 & 0.21 $\pm$ 0.03 \\ 
      & 12.38 & H I + He I   & 12.37 + 12.39 & 1.36 $\pm$ 0.05 \\ 
      & 16.22 & H I + He I   & 16.21 + 16.20 & 0.71 $\pm$ 0.04 \\ 
      & 19.07 & H I + He I   & 19.06 + 19.05 & 1.45 $\pm$ 0.22 \\ 
      & 22.35 & H I + He I   & 22.33 + 22.32 & 2.21 $\pm$ 0.13 \\ 
      & 27.82 & H I + He I   & 27.80 + 27.79 & 1.96 $\pm$ 0.03 \\
      & 29.84 & H I          & 29.84         & 1.10 $\pm$ 0.10 \\ \hline
      HD~209975 &  &  &  &    \\ 
      & 11.32 & H I + He I    & 11.31 + 11.30 & 0.17 $\pm$ 0.02 \\ 
      & 12.38 & H I + He I    & 12.37 + 12.39 & 1.02 $\pm$ 0.04 \\ 
      & 16.22 & H I + He I    & 16.21 + 16.20 & 0.80 $\pm$ 0.02 \\ 
      & 19.07 & H I + He I    & 19.06 + 19.05 & 1.05 $\pm$ 0.04 \\ 
      & 22.35 & H I + He I    & 22.33 + 22.32 & 1.66 $\pm$ 0.22 \\ 
      & 27.82 & H I + He I    & 27.80 + 27.79 & 1.01 $\pm$ 0.29 \\ \hline
      HD~195592 &  &  &  &    \\ 
      & 10.51 & H I + He I  & 10.50 + 10.50 &  1.06 $\pm$ 0.03 \\ 
      & 11.32 & H I + He I  & 11.31 + 11.30 &  0.84 $\pm$ 0.01 \\ 
      & 12.38 & H I + He I  & 12.37 + 12.39 &  4.06 $\pm$ 0.05 \\ 
      & 16.22 & H I + He I  & 16.21 + 16.20 &  1.33 $\pm$ 0.05 \\ 
      & 16.88 & H I + He I  & 16.88 + 16.87 &  0.97 $\pm$ 0.01 \\ 
      & 19.07 & H I + He I  & 19.06 + 19.05 &  4.83 $\pm$ 0.23 \\ 
      & 22.35 & H I + He I  & 22.33 + 22.32 &  2.80 $\pm$ 0.05 \\ 
      & 27.82 & H I + He I & 27.80 + 27.79  &  8.42 $\pm$ 0.41 \\
      & 29.84 & H I        & 29.84          &  5.10 $\pm$ 0.30 \\ 
      & 30.02 & H I        & 30.02          &  3.00 $\pm$ 0.30 \\ \hline \hline
    \end{tabular}
  \end{table*}
\end{center}

\begin{figure*} 
\centering
 \makebox[\textwidth]{\includegraphics[trim= 0mm 0mm 0mm 10mm,height=.60\paperheight,width=1.05\paperwidth,angle=180]{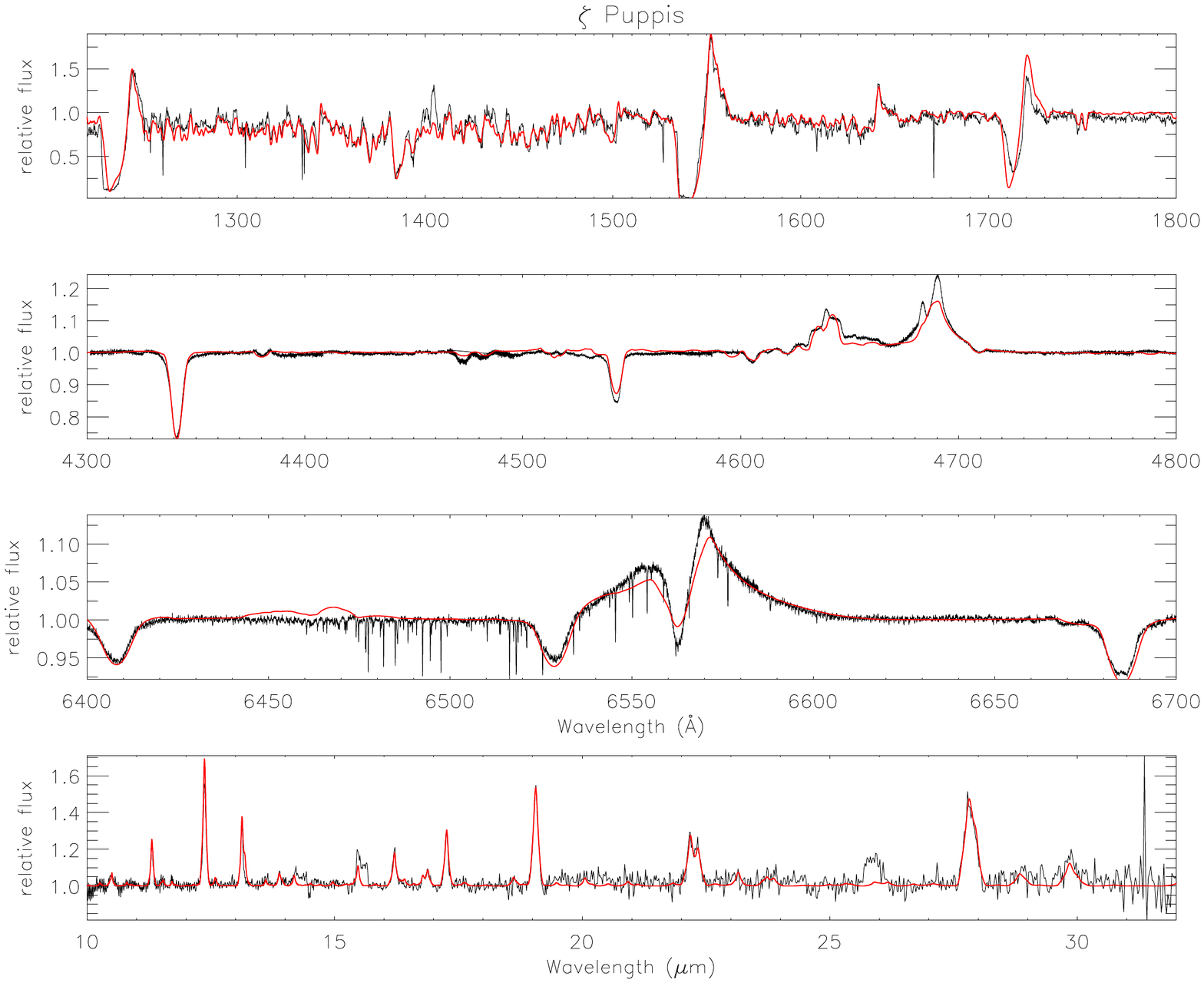}}
\caption{Model fits (red; thick line) to the UV (IUE), optical (FEROS)
and mid-IR (Spitzer) data of $\zeta$ Puppis.}
\label{finalzeta}
\end{figure*}

\begin{figure*}
\centering
\includegraphics[trim= 20mm 20mm 20mm 20mm,width=15.0cm,height=11.0cm]{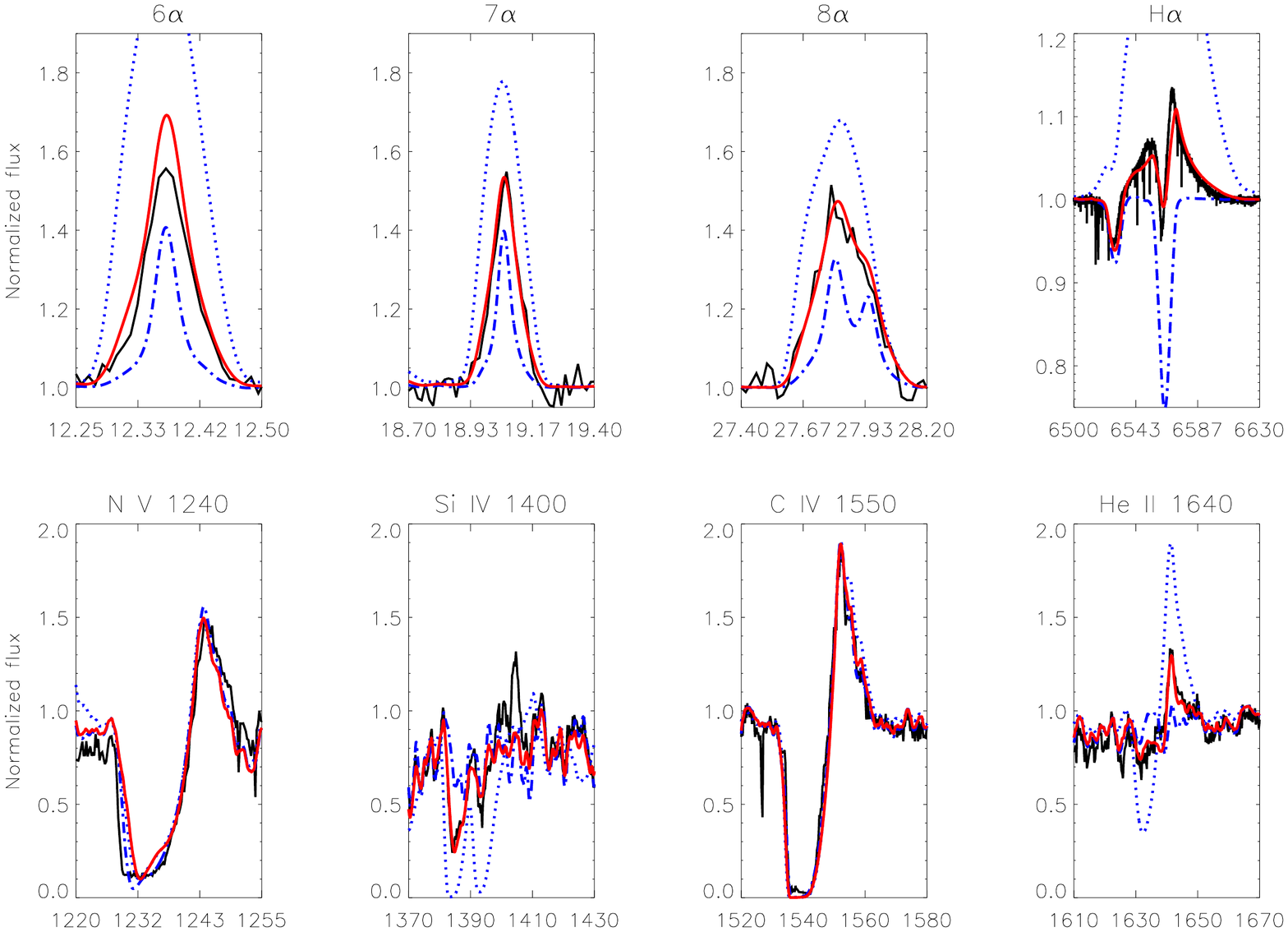}
\caption{Sensitivity of mid-IR lines to mass-loss rate changes (wavelength given in $\mu$m).
The chosen star is $\zeta$ Puppis. Selected UV lines and H$\alpha$ are
also shown (wavelengths in \AA). The rates displayed are: $7.0 \times 10^{-7}$
M$_{\sun}$ yr$^{-1}$ (blue dashed-dotted line),  $1.9 \times 10^{-6}$ M$_{\sun}$ yr$^{-1}$ (final model;
red solid line), and $6.0 \times 10^{-6}$ M$_{\sun}$ yr$^{-1}$ (blue dotted line).
A unique model spectrum with the same mass-loss rate matches the mid-IR, optical and UV observations (black).}
\label{mdots}
\end{figure*}

\begin{figure*} 
\centering
\includegraphics[trim= 20mm 0mm 40mm 10mm,angle=180,scale=0.85]{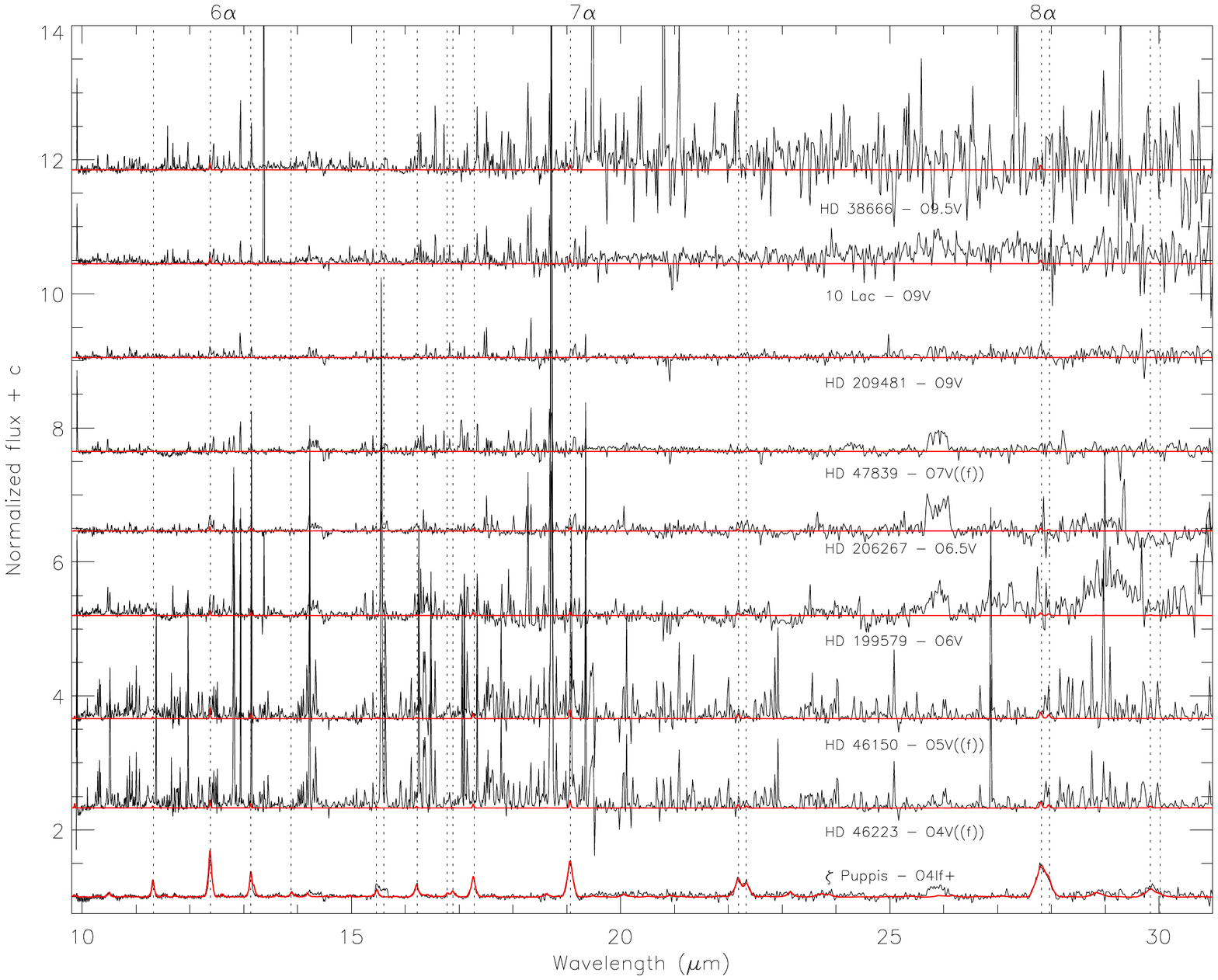}
\caption{The mid-IR {\it Spitzer} spectra of the dwarfs of our sample (black
lines). For comparison, the bottom panel show the observed and synthetic spectra
of $\zeta$ Puppis. From bottom to top we have early to late spectral types.
Synthetic CMFGEN spectra are displayed in red. They support the conclusion that
the mid-IR spectrum of O dwarfs does not contain conspicuous atmospheric features.}
\label{dwfs}
\end{figure*}


\bsp	
\label{lastpage}
\end{document}